\documentclass[seceq,supplement]{ptptex}
\usepackage{graphicx}

\def\figp#1#2#3#4{
\begin{figure}[!tp]
\begin{center}
\includegraphics[width=#3\textwidth,bb=#4]{#1.pdf}
\caption{#2}
\label{fig:#1}
\end{center}
\end{figure}}
\def\muf{{\mu_{\rm F}}}
\def\muw{{\mu_{\rm W}}}
\def\mus{{\mu_{\rm S}}}

\def\pf{p^{(\rm F)}(c)}
\def\pfc{P^{(\rm F)}_>(c)}
\def\pw{p^{(\rm W)}(c)}

\def\sumk{\sum_{k=1}^K}
\def\pk{p_k}
\def\act{\langle c \rangle_{\beta}}
\def\actn#1{\langle c^#1 \rangle_{\beta}}
\def\ac{\langle c \rangle_{0}}
\def\acn#1{\langle c^#1 \rangle_{0}}

\def\pb{p^{\rm (GB2)}(c)}
\def\xx{\biggl(\frac{c}{c_1}\biggr)}
\def\xxi{\biggl(\frac{c_1}{c}\biggr)}
\title{%
Labour Productivity Superstatistics\footnote{Talked presented at 
the Yukawa Institute workshop {\it ``Econophysics III 
--Physical approach to social and economic phenomena"}
(YITP-W-07-16) on Dec.24, 2007.}}%

\author{
Hideaki \textsc{Aoyama}$^{1,}$\footnote{hideaki.aoyama@scphys.kyoto-u.ac.jp},
Hiroshi \textsc{Yoshikawa}$^2$,
Hiroshi \textsc{Iyetomi}$^3$,
Yoshi \textsc{Fujiwara}$^4$
}

\inst{
$^1$Physics Department, Kyoto University, Kyoto 606-8501, Japan\\
$^2$Faculty of Economics, University of Tokyo, Tokyo 113-0033, Japan\\
$^3$Department of Physics, Niigata University, Niigata 950-2181, Japan\\
$^4$NiCT/ATR CIS, Applied Network Science Lab., Kyoto 619-0288, Japan
}

\abst{
We discuss superstatistics theory of labour productivity.
Productivity distribution across workers, firms and 
industrial sectors are studied empirically and found to
obey power-distributions, in sharp contrast to
the equilibrium theories of mainstream economics.
The Pareto index is found to decrease with the level of aggregation,
{\it i.e.}, from workers to firms and to industrial sectors.
In order to explain these phenomenological laws, we propose a
superstatistics framework, where the role of the
fluctuating temperature is played by the fluctuating demand.}

\begin{document}

\maketitle
\centerline{KUNS-2140}

\section{Introduction}
In mainstream macroeconomics, the marginal productivity is believed to be equal
across workers, firms, and business sectors,
which is identified with nature of equilibrium:
If the productivity is not equal, there will be profit-opportunity, 
which, according to the standard economic equilibrium theory, is against the notion of equilibrium.
This, however, is far from physics understanding:
When a system made of many constituents, such as 
gas made of molecules, is in an equilibrium state, physical quantities, such as
energy of the constituents is not unique, but is distributed
obeying certain statistical laws, such as Boltzmann law.
The exchange of energy between the constituents do not equalize the energy, but
manifests itself in the realization of the statistical distribution.
Keynes' economics \cite{Keynes} is somewhat close to our physics view, allowing
involuntary unemployment, which leads to distributions, although
wider perspective may be desired.

Thus, the study of the distribution of the productivity\footnote{In stead of 
``distribution", the term ``dispersion" is commonly used in Economics.
Thus, ``productivity dispersion" in stead of ``productivity distribution".
Since this conflicts with normal physics usage of ``dispersion", we will
use the latter word for accuracy.} is
an important issue in establishing the notion of the economic equilibrium.

We approach this problem just as in any physics, or rather, any branch of science:
We fist study the phenomena itself to find phenomenological 
laws,\footnote{``Stylized facts" in Economics terminology.}
and then will search for theoretical understanding behind them,
seeking further tests and refinements.

\section{Phenomenology}
The database we have used is the Nikkei-NEEDS 
(Nikkei Economic Electronic Databank System) database\cite{nikkei},
which contains financial data of all the listed firms in Japan.
As such, it is a well-established and representative database, 
widely used for various purposes from research to practical business applications.
For our purpose, we used their 2007 CD-ROM version and extracted data for the period 
between 1980 and 2006, which contained 
some 1,700 to 3,000 firms and 4 to 6 million workers, 
with the numbers increasing with the year.

In the following, we study the labour productivity $c$;
\begin{equation}
c:=\frac{Y}{L}\,,
\end{equation}
where $Y$ is the production (in currency) and $L$ is the labour (in number of workers).
To be exact, this is the {\it mean} labour productivity, different from the
{\it marginal} labour productivity $\partial Y /\partial L$
dealt in the equilibrium theory of the economics.
This difference, however, does not affect the following discussion of the
superstatistics theory,\cite{ayif} due to the fact that they obey Pareto distribution
as we will see below.
Also, we  have found that the values of $c$ 
calculated from this database are sometimes inappropriately large.
Part of this may come from the fact that ``the number of employees"
reported in this database does not contain temporally workers.
Also, firms that became stock-holding firms reported huge reduction
of the number of employees (as it is defined to be the value at the
{\it end} of that year), while maintaining same order of sales revenue 
in the year the conversion occurred. This results in
absurd values of the productivity $c$ for that year.
Because of these abnormalities, we have excluded top-ten firms
in terms of the productivity each year.
This roughly corresponds to excluding firms with productivity $c > 10^9$ yen/person.
(We have carried out analysis with several different cuts, {\it i.e.},
with cutting top-twenty firms, cutting on the value of $c$, etc.), but
the result remained mostly stable.)

\figp{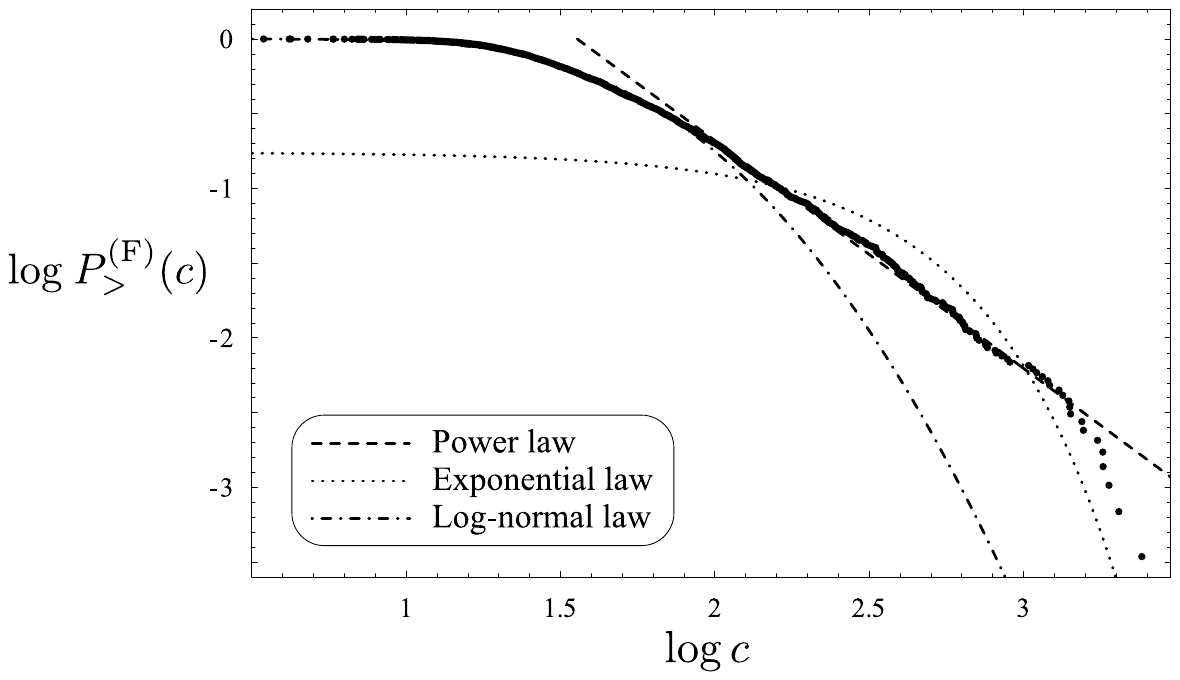}{\textsc{Productivity Distribution across Firms (2005)} 
{\it Notes:} The productivity $c$ is in the unit of $10^6$yen/person.
The best fits for the exponential law and the power law
is obtained for $10<c<3000$.}{0.7}{220 192 617 393}

\figp{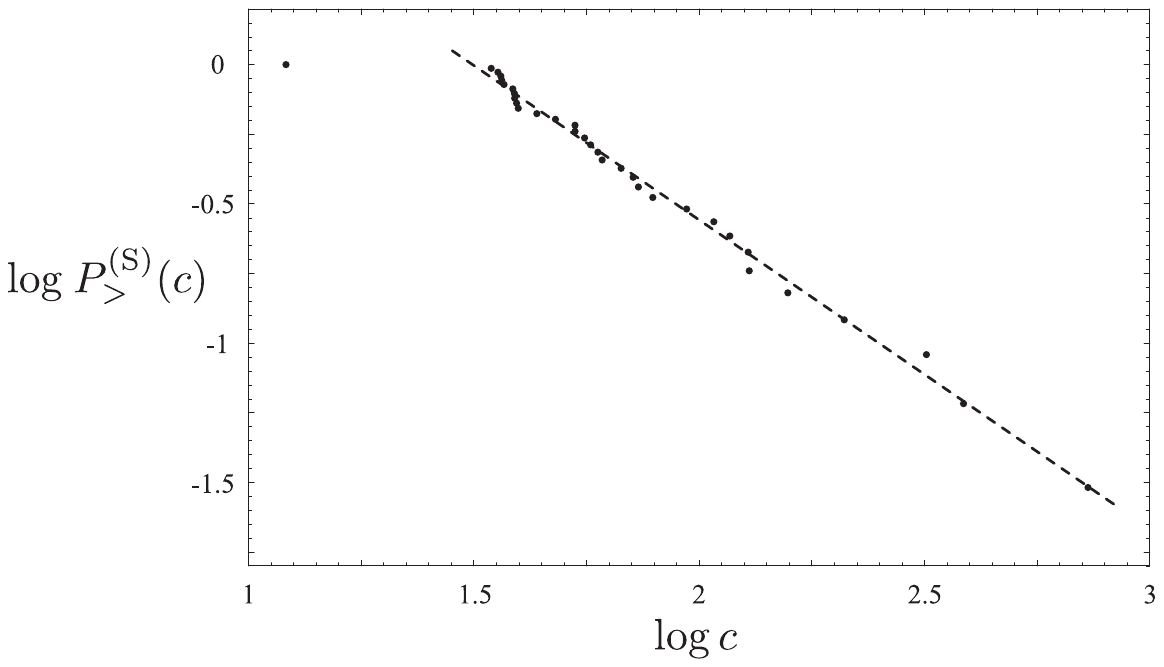}{\textsc{Productivity Distribution across 
{\it Business Sectors} (2005)}}{0.7}{219 198 616 399}

\figp{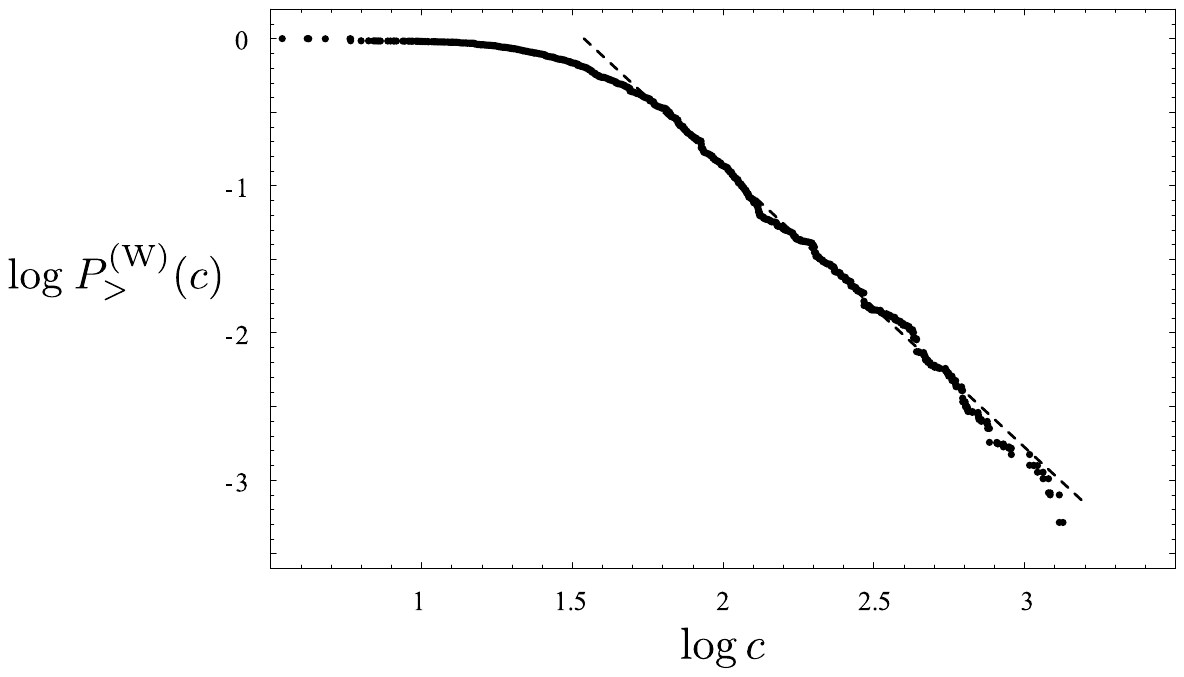}{\textsc{Productivity Distribution across {\it Workers} (2005)}}
{0.7}{422 190 819 391}

The plots of the cumulative distribution function (cdf) of the productivity,
which is defined in terms of the probability density function (pdf) $p(c)$ as;
\begin{equation}
P_>(c):=\int_c^\infty p(c) \,dc,
\end{equation}
are given
in Figs.\ref{fig:firm2005_Nikkei}--\ref{fig:employee2005_Nikkei} 
for three different levels of aggregation in 2005 \cite{ayif}.

As is evident from these figures, the productivity $c$ obeys the Pareto law (power-law) in
the asymptotic region at all three levels:
\begin{equation}
p_>(c)\propto c^{-\mu}
\quad (c\rightarrow \infty),
\label{paretodef}
\end{equation}
The exponent $\mu$ is called Pareto index.
This qualitative feature is true for all the years we have studied. 

Study of the Pareto index $\mu$ calls for a careful analysis:
Since the power-law is a straight line in the log-log plot of the
cdf, it is tempting, at least for a novice,  to take the
(almost-)straight section of the plot and fit is linearly, obtaining
the Pareto index as the gradient of the best-fit straight line.
This is rather dangerous, as 
its value often depends on the choice of the section, which is somewhat arbitrary.
Even if a definition of the section is good for a particular year,
it may become inappropriate in later years.
Changing the choice of the section from a year to another year
will destroy objectivity of discussion of the evolution of the Pareto index over the years.

\def\pb{p^{\rm (GB2)}(c)}
\def\xx{\biggl(\frac{c}{c_1}\biggr)}
\def\xxi{\biggl(\frac{c_1}{c}\biggr)}
For these reasons, it is desirable to use a distribution defined for the whole
region of $c\in [0, \infty)$ that behaves as a power-law in the asymptotic region.
One such distribution that is suitable for numerical analysis 
and is general enough for our purpose is the 
``Generalized Beta Distribution of the Second Kind'' (GB2)  \cite{actuarial}.
It is defined by the following cdf:
\begin{equation}
P^{\rm (GB2)}_>(c)=\frac{B(z,\mu/q,\nu/q)}{B(\mu/q,\nu/q)},
\quad
z=\left[\,1+\xx^q\right]^{-1}. 
\end{equation}
where $\mu,\nu,q,c_1>0$ and 
$B(z,s,t)$ is the incomplete Beta function
with $B(1,s,t)=B(s,t)$.
The corresponding pdf is the following:
\begin{align}
\pb
&=\frac{q}{B(\mu/q,\nu/q)}
\frac1c \xx^{\nu}\left[\,1+\xx^q\right]^{-(\mu+\nu)/q}.
\label{bb2}
\end{align}
For large $c$ this behaves as follows 
\begin{equation}
\pb\simeq \frac{q}{B(\mu/q,\nu/q)}\frac1c\xx^{-\mu},
\end{equation}
and its cdf as;
\begin{equation}
P_>^{\rm (GB2)}(c)\simeq \frac{q/\mu}{B(\mu/q,\nu/q)}\xx^{-\mu}.
\end{equation}
Therefore, the parameter $\mu$ is the Pareto index and
the parameter $c_1$ a scale for the power law.
For $c\rightarrow 0$, this distribution also reduces to a power law
with the power exponent equal to $\nu$.

In general, the parameter $q$ determines how persistent two power-laws 
at both ends are; for small $q$ the transition from the small--$c$ power law
to the large--$c$ power law is smooth. In such a case, 
it can be approximated by the log-normal distribution around its peak at 
\begin{equation}
c_{\ln}=\left(\frac{\nu}{\mu}\right)^{1/q}
\end{equation} 
and 
\begin{equation}
\sigma=\frac1q\left(\frac{\nu+\mu}{\nu\mu}\right)\,.
\end{equation}
As the log-normal distribution is widely observed for small-to-medium range in real 
economic distributions, this model-distribution has a good chance of being a valid approximation.

\figp{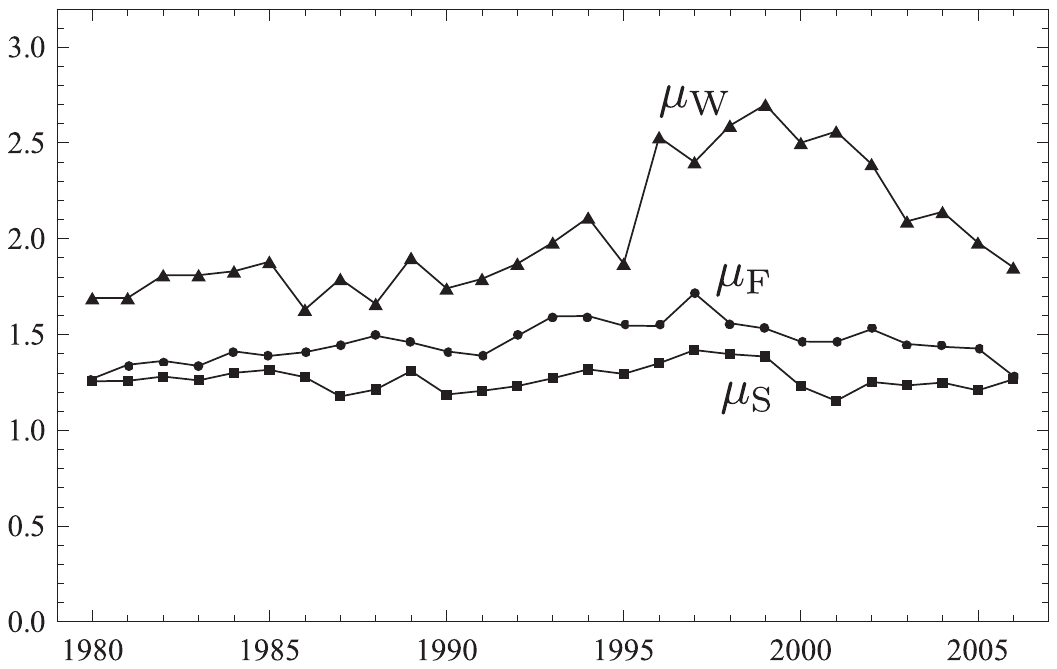}{\textsc{Pareto Indices of the Productivity Distributions 
across Workers, Firms, and Industrial Sectors}}{0.6}{147 325 447 515}
Fitting the data with this GB2 distribution by the maximum likelihood method, 
we have obtained the values of Pareto index plotted in Fig.\ref{fig:ParetoIndex}.
From these, we find the following two phenomenological laws.

\vspace{10pt}

\begin{enumerate}
\item[I.]
The distribution of productivity obeys the Pareto distribution
(\textit{i.e.} the power-law for the high productivity group)
at every level of aggregation, that is, across workers, firms, and industrial sectors.
\item[II.]
The Pareto index, namely the power exponent decreases as the level of aggregation goes up: $\mu_W > \mu _F > \mu _S$.
\end{enumerate}

\vspace{10pt}\noindent
In the following, we present theoretical framework to explain these laws.

\section{Yoshikawa-Aoki theory}
Yoshikawa and Aoki\cite{yoshikawa, aokiyoshikawa} proposed an equilibrium theory of productivity
distribution several years ago. The key ideas are summarized 
by the following correspondence with the statistical physics:

\begin{align}
\textbf{Economics} &\Leftrightarrow  \textbf{Physics} \nonumber \\
\textrm{Firms} &\Leftrightarrow  \textrm{Energy Levels} \nonumber\\
\textrm{Workers} &\Leftrightarrow  \textrm{Molecules} \nonumber\\
\textrm{Worker's Productivity} &\Leftrightarrow  \textrm{Molecule's Energy} \nonumber\\
\textrm{Aggregate Demand} &\Leftrightarrow  \textrm{Total Energy} \nonumber
\end{align}

The last correspondence may be explained as follows:
Let us denote the total number of firms by $K$ and each firm
is labeled by an index $k$ ($k=1,2,\dots,K$).
The number of workers $n_k$ at the firm $k$ is
constrained by the total number of workers $N$;
\begin{equation}
\sumk n_k = N.
\label{ncons}
\end{equation}
The productivity $c_k$ is constrained by the fact that
the sum of firm's production is the total production, which is equal to the aggregate 
demand $\tilde{D}$;
\begin{equation}
\sumk n_k c_k =\tilde{D}.
\label{dcons}
\end{equation}
This constraint (\ref{ncons}) corresponds to that on the total number
of molecules, while (\ref{dcons}) to that on the total energy.

Yoshikawa and Aoki postulated that the actual distribution is the
one that maximized the entropy, arriving at the Boltzmann law for
the probability $\pk$ of the worker's productivity being equal to $c_k$;
\begin{equation}
\pk:=\frac{\langle n_k\rangle}{N}=\frac1{Z(\beta)} \, e^{-\beta c_k}.
\label{boltz}
\end{equation}
where $Z(\beta)$ is the usual partition function:
\begin{equation}
Z(\beta):= \sumk e^{-\beta c_k}.
\label{partid}
\end{equation}
The inverse-temperature $\beta$ is determined by the mean demand $D$ as follows:
\begin{equation}
D:=\frac{\tilde{D}}{N}=-\frac{d}{d\beta}\ln Z(\beta).
\label{dnbeta}
\end{equation}

The above result shows that the productivity distribution of workers
is determined by (1) the productivity distribution of firms and (2) the 
mean demand $D$. This is readily exposed in the continuous notation:
Denoting the pdf of the firm's productivity 
by $\pf$, the pdf of the worker's productivity $\pw$ is given as follows;
\begin{equation}
\pw=\frac1{Z(\beta)} \, e^{-\beta c} \pf,
\label{pee}
\end{equation}
where the partition function is 
\begin{equation}
Z(\beta):=\int_o^\infty e^{-\beta c} \pf dc.
\label{parti}
\end{equation}
The firm's productivity distribution $\pf$ is analogous to the density of energy levels.

Although this theory is rather attractive for its simplicity and elegance, 
it does not meet either of the phenomenological laws I and II we have established above:
Given that $\pf$ obeys power-law, this theory predicts that $\pw$ has additional
exponential dumping, which is far from the reality.
We thus proceed to superstatistics theory build on this platform.

\section{Fluctuating Aggregate Demand and Superstatistics}

The implicit assumption in the Yoshikawa-Aoki theory that the aggregate demand $\tilde{D}$ 
is constant is an oversimplification; the demand fluctuates. 

Macro-system under fluctuations of
external environment may be handled by the \textit{superstatistics} \cite{beckcohen,beck1}.
In this theory, the system goes through changing external influences, but is in equilibrium 
within certain limited scale in time and/or space, in 
which the temperature may be regarded as constant and
the Boltzmann distribution is achieved.
In other words, the system is only locally in equilibrium;
globally seen, it is out of equilibrium.
Thus the key concept in the superstatistics is to 
introduce averaging over the Boltzmann factors. 

There are several model-cases where superstatistics was
applied successfully.
Among them,  the Brownian motion of a particle going thorough changing
environment\cite{ausloos2006bph,lz}, such as different temperature
and different viscosity both in space and time, 
provides a good analogy to our case.
Like a particle in Brownian motion, 
each firm experiences ever-changing demand, the first of which is the aggregate demand $\tilde{D}$.
Further, firms in different business sector or with different 
product must meet their own and individual demand, which is analogous to the
special fluctuation of environment.
Average over these various possible fluctuations could be summarized in
a form of averaging over the temperature, or in other words, $D$.
Although it is dangerous to extend this interpretation too far with a lack of
rigorous treatment of the associating stochastic process,
we would always keep in mind that the following discussion of the
fluctuation of the aggregate demand may be taken as a symbolic representation
of many other kinds of fluctuation.

In superstatistics, the familiar Boltzmann factor $e^{-\beta  c}$ is replaced by
a weighted average:
\begin{equation}
B(c)=\int_0^\infty e^{-\beta c} f_\beta(\beta) d\beta,
\label{bc}
\end{equation}
Here, the weight factor $f_\beta(\beta)$ represents the changing environment.
Note that because $\beta $ is a monotonically decreasing function of the mean demand $D$, 
the weight factor $f_\beta(\beta)$ represents fluctuation of $D$.
With this weight factor, the pdf of worker's productivity
(\ref{pee}) is now replaced by the following:
\begin{equation}
\pw = \frac1{Z_B} \pf B(c).
\label{peess}
\end{equation}
Here, the partition function $Z_B$ is also redefined as 
\begin{equation}
Z_B=\int_0^\infty \pf B(c) dc.
\label{zbpss}
\end{equation}

Let us now examine whether $\pw$ in Eq.(\ref{peess}) obeys the
Pareto law for high productivity $c$.
As the integration in Eq.(\ref{bc}) is dominated by the small $\beta$
(high demand) region for large  $c$,
the behaviour of the pdf $f_\beta(\beta)$ for small $\beta$ is critical.
Let us assume the following in this range:
\begin{equation}
f_\beta(\beta) \propto \beta^{-\gamma}, \quad (\gamma<1),
\label{fbeta}
\end{equation}
where the constraint for the parameter $\gamma$ 
comes from the convergence of the integration in Eq.(\ref{bc}).
The proportional constant is irrelevant because $\pw$ is normalized by $Z_B$.
This leads to the following  $B(c)$ for large $c$:
\begin{equation}
B(c) \propto   \Gamma(1-\gamma)\, c^{\gamma-1}.
\label{bcgamma}
\end{equation}
Substituting this and the Pareto law for the firm's productivity,
\begin{equation}
\pf \propto c^{-\muf-1}
\label{pfasym}
\end{equation}
into Eq.(\ref{peess}), we obtain
the productivity distribution across workers obeys the Pareto law;
\begin{equation}
\pw \propto c^{-\muw-1},
\label{mue}
\end{equation}
with
\begin{equation}
\muw=\muf-\gamma+1.
\label{muef1}
\end{equation}
Because of the constraint $\gamma <1$, this leads to 
the inequality
\begin{equation}
\muw > \muf.
\end{equation}
This agrees with our empirical observation II.

The parameter $\gamma$ in the distribution of $\beta$ is related to
a parameter in distribution of $D$, as $\beta$ is related to $D$ by Eq.(\ref{dnbeta}).
Therefore, let us now examine the consequence of the relation (\ref{dnbeta}) for 
small $\beta$.

\section{$\beta$ and $D$}
We first note the following three basic properties (i)--(iii).

\vspace{10pt}
\begin{enumerate}
\item[(i)]
The temperature, $T=1/\beta$ is a monotonically increasing function of the aggregate demand, $D$.
We can prove it using Eq.(\ref{dnbeta}) as follows:
\begin{align}
\frac{dD}{dT}=
-\frac1{T^2}\frac{dD}{d\beta}
=\beta^2\frac{d^2}{d\beta^2}\ln Z(\beta)
=\beta^2\left(\actn{2}-\act^2\right) \ge 0.
\end{align}
where $\actn{n}$ is the $n$-th moment of productivity defined as follows:
\begin{equation}
\actn{n}\equiv\frac1{Z(\beta)}\int_0^\infty c^n \pf \,e^{-\beta c}\,dc.
\label{momdef}
\end{equation}
Note that $\act=D$.
This is a natural result. As the aggregate demand $D$ rises, 
workers move to firms with higher productivity.
It corresponds to the higher temperature due to the weight factor $e^{-\beta c}$.
\item[(ii)]
For $T\rightarrow 0$ ($\beta\rightarrow\infty$),
\begin{equation}
D\rightarrow 0.
\end{equation}
This is evident from the fact that in the same limit the integration
in Eq.(\ref{parti}) is dominated by $c\simeq 0$ due to the factor $e^{-\beta c}$.
\item[(iii)]
For $T\rightarrow \infty$ ($\beta\rightarrow 0$),
\begin{equation}
D\rightarrow \int_0^\infty c\, \pf \,dc\ (=\ac). 
\end{equation}
This can be established based on the property (i) because 
$D=\act\rightarrow \ac$ as $\beta\rightarrow 0$ and 
$Z(0)=1$.
\end{enumerate}

\vspace{10pt}

Let us now study the small $\beta$ (high temperature) properties.
One possible approximation for Eq.(\ref{parti}) is obtained by expanding
the factor $e^{-\beta c}=1-\beta c +\cdots$ and carrying out the
$c$-integration in each term. This leads to the following:
\begin{align}
Z(\beta)&= \int_0^\infty \pf \left(1-\beta c +\frac12(\beta c)^2+\dots 
\right) dc\nonumber\\
&= 1 - \ac\beta + \frac12 \acn{2}\beta^2+ \dots,
\label{notright}
\end{align}
where we have used the normalization condition,
\begin{equation}
\int_0^\infty\pf\, dc =1.
\end{equation}
The result (\ref{notright}) is, however, valid only for $\muf>2$
since $\acn{2}$ is infinite for $\muf \le 2$, 
which is true as we have seen.

The correct expansion for  $1< \muf < 2$ is done in the following way.
We first separate out the first two terms in the expansion of the 
factor $e^{-\beta c}$;
\begin{align}
Z(\beta)&=\int_0^\infty \pf \left(1-\beta c +(e^{-\beta c}-1+\beta c)\right)dc
\nonumber\\
&=1-\ac \beta +Z_2(\beta),\\
Z_2(\beta)&\equiv\int_0^\infty \pf \,g(c)dc
=\int_0^\infty \left(-\frac{\partial}{\partial c} \pfc\right) g(c)dc
\nonumber\\
&=\int_0^\infty \pfc \frac{\partial g(c)}{\partial c}dc,
\label{z2good}
\end{align}
where $g(c)=e^{-\beta c}-1+\beta c$ is a monotonically increasing function of $c$ with 
\begin{equation}
g(0)=g'(0)=0.
\label{g00}
\end{equation}
The $c$-integration in Eq.(\ref{z2good}) is dominated by the asymptotic region of $c$
for small $\beta$. Therefore, the leading term in $Z_2(\beta)$ is evaluated by
substituting the asymptotic expression of $\pf$;
\begin{equation}
\pfc \simeq \left(\frac{c}{c_0}\right)^{-\muf},
\end{equation}
into Eq.(\ref{z2good}). We thus arrive at the following:
\begin{align}
Z_2(\beta)&=\int_0^\infty \left(\frac{c}{c_0}\right)^{-\muf} \frac{\partial g(c)}{\partial c}dc
+\cdots
\nonumber\\
&=\muf\Gamma(-\muf)(c_0\beta)^\muf+\cdots.
\end{align}

The case $\muf=2$ can be obtained by taking the limit
$\muf\rightarrow 2+$ in the following expansion valid for $2<\muf<3$:
\begin{equation}
Z(\beta)
= 1 - \ac\beta + \frac12 (\acn{2}-\ac^2)\beta^2
+ \muf\Gamma(-\muf) (c_0 \beta)^\muf+\dots.
\end{equation}
which can be obtained in the manner similar to the above.
The third term is finite for $\muf>2$, but
diverges as $\muf\rightarrow 2+$ as
\begin{equation}
\acn{2}\rightarrow \frac{2c_0^2}{\muf-2}.
\end{equation}
This cancels the divergence of the fourth term in the same limit
and the remaining leading term is as follows:
\begin{equation}
Z(\beta)
= 1 - \ac\beta  - (c_0 \beta)^2 \log(c_0\beta)+\dots.
\end{equation}

In summary, the partition function behaves as follows:
\begin{equation}
Z(\beta)=
\begin{cases}
1- \ac\beta + \frac12 \acn{2}\beta^2+ \dots
& \mbox{for } 2<\muf;\\
1- \ac\beta-(c_0\beta)^2\log(c_0\beta)+\dots
& \mbox{for } \muf=2;\\
1- \ac\beta+ \muf\Gamma(-\muf) (c_0 \beta)^\muf+\dots
& \mbox{for } 1<\muf<2.\\
\end{cases}
\end{equation}
Substituting the above in Eq.(\ref{dnbeta}), we obtain the following:
\begin{equation}
D=
\begin{cases}
\ac-\left(\acn{2}-\ac^2\right)\beta+\dots.
& \mbox{for } 2<\muf;\\
\ac+2c_0^2\beta\log(c_0\beta)+\dots
& \mbox{for } \muf=2;\\
\ac - \muf^2 \Gamma(-\muf) c_0^\muf \beta^{\muf-1} +\dots
& \mbox{for } 1<\muf<2.\\
\end{cases}
\label{hight}
\end{equation}

\section{Pareto Indices and the Demand}

As the distributions of $\beta$ and $D$ are related by
\begin{equation}
f_\beta (\beta)d\beta=f_{D}(D)dD,
\end{equation}
we find from Eqs.(\ref{fbeta}) and (\ref{hight}) that
\begin{equation}
f_{D}(D)\propto\left(\ac-D \right)^{-\delta},
\label{deltadef}
\end{equation}
with 
\begin{equation}
\gamma-1=
\begin{cases}
\delta-1 
&\mbox{for }2<\muf;\\
(\muf-1)(\delta-1)
&\mbox{for }1<\muf<2.
\end{cases}
\label{gammarho}
\end{equation}
At $\muf=2$, we need additional logarithmic 
factors for $f_\beta(\beta)$, but the power of $\beta$ is essentially 
the boundary case between the above two, $\gamma=\delta$.
Also, the parameter $\delta$ is constrained by;
\begin{equation}
\delta<1
\end{equation}
from the normalizability of the distribution of $f_{D}(D)$,
which is consistent with the constraint $\gamma<1$ and Eq.(\ref{gammarho}).

Combining Eqs.(\ref{muef1}) and (\ref{gammarho}),
we reach the following relation between the Pareto indices:
\begin{equation}
\muw= \begin{cases}
\muf-\delta+1 & \mbox{for } 2<\muf \\
(\muf-1)(-\delta+1)+\muf &\mbox{for } 1<\muf<2.
\end{cases}
\label{mumu}
\end{equation}
This relation between $\mu _W$ and $\mu _F$ 
is illustrated in Fig.\ref{fig:mumu}.
As noted previously, because of the constraint $\delta<1$, 
Eq.(\ref{mumu}) necessarily makes $\muw$ larger than $\muf$,
in good agreement with our empirical finding. 
Incidentally, Eq.(\ref{mumu}) has a fixed point at $(\muw,\muf)=(1,1)$;
the line defined by Eq.(\ref{mumu}) always passes through this point irrespective of
the value of $\delta$. 
The Pareto index for firms is smaller than that for workers, but it
 cannot be less than one, because of the existence of this fixed point.

\figp{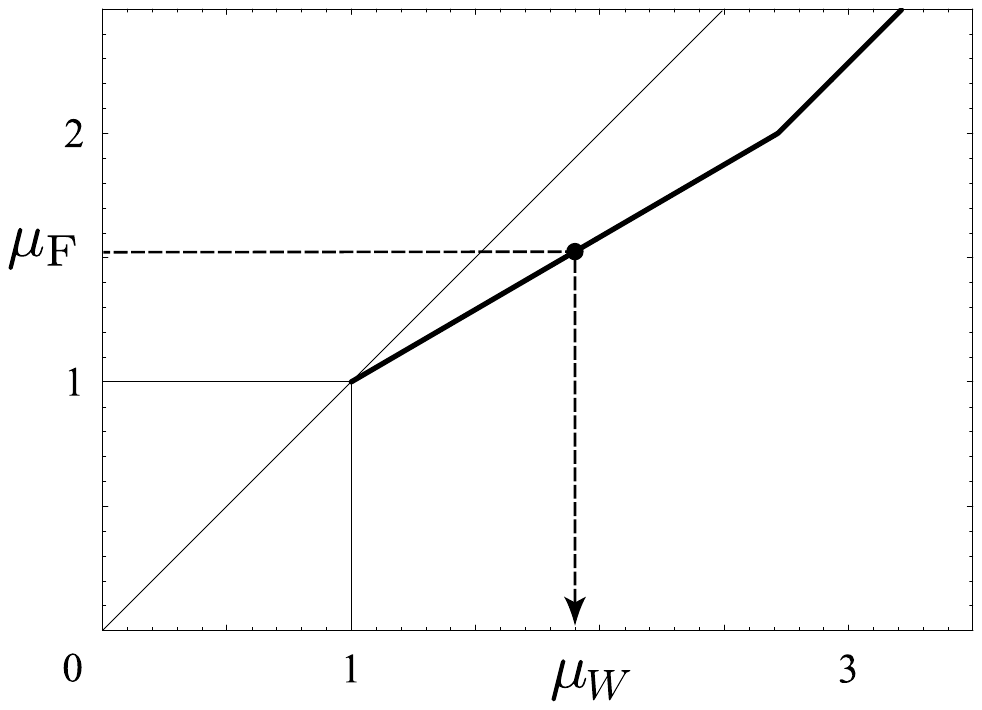}{Illustration of the relation between $\muw$ and $\muf$ (\ref{mumu}). 
The solid line is the relation (\ref{mumu}), and the filled circle 
is the data.}{0.5}{138 413 415 613}

The superstatistics framework presented above may apply
for any adjoining levels of aggregation;
Instead of applying it for workers and firms, we may 
apply it for firms and industrial sectors.
Then, we can draw the conclusion
 that as we go up from firms to industrial sectors, the
Pareto index again goes down, albeit for a  different value of 
$\delta$.
This is illustrated in Fig.\ref{fig:mumumu}.
Because of the existence of the point (1, 1), 
 as the aggregation level goes up,
the Pareto index is driven toward 1, but not beyond 1.
{\it At the highest
aggregation level, it is expected to be close to one}.
This is again in good agreement with our empirical finding that
the Pareto index of the industrial sector $\mus$ is 
 close to one (see Fig.\ref{fig:ParetoIndex}).

\figp{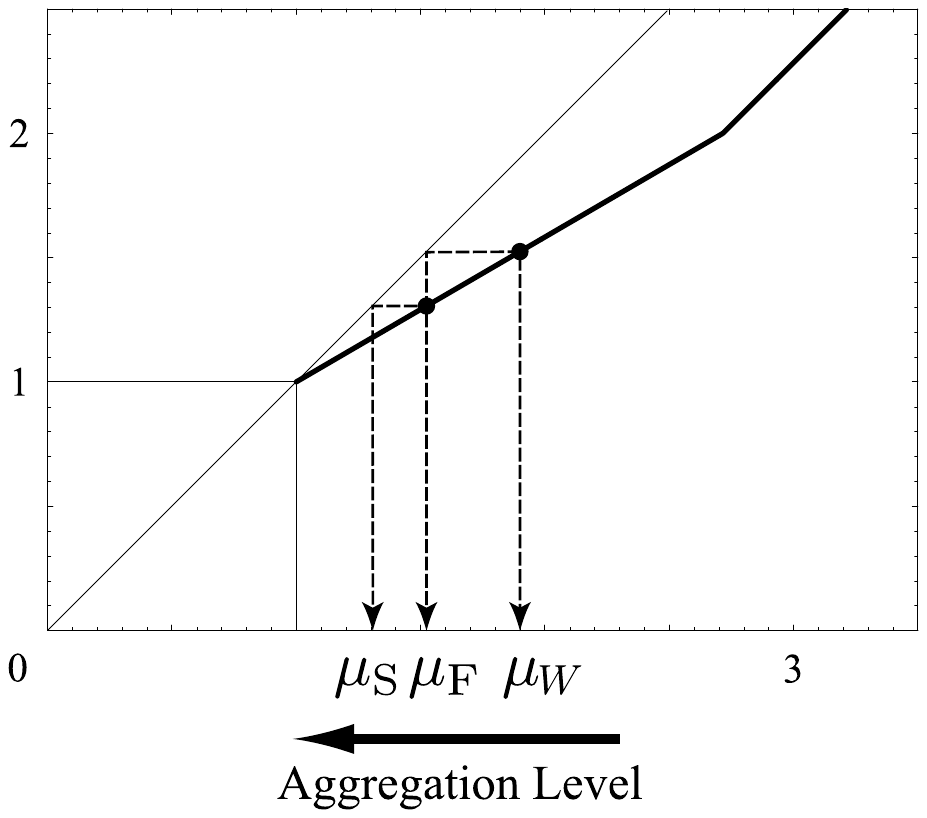}{Illustration of changes of the Pareto index
as the aggregation level changes in two steps, each with 
a different value of $\delta$.}{0.47}{153 382 415 613}

In summary, the superstatistics framework successfully explains two empirical findings
we have made.
Furthermore, given the measured values of $\muw$ and $\muf$,
the relation (\ref{mumu}) can be used to determine the value of $\delta$:
\begin{equation}
\delta=
\begin{cases}
\muf-\muw+1 & \mbox{for } 2<\muf;\\[5pt]
\displaystyle \frac{\muf-\muw}{\muf-1}+1 &\mbox{for } 1<\muf<2.
\end{cases}
\label{deltais}
\end{equation}
The result is shown in Fig.\ref{fig:delta}.
Recall that $\delta $ is the power exponent of the distribution of aggregate demand, $D$.  
Therefore, 
low $\delta$ means the relatively low level of the aggregate demand. 
In Fig.\ref{fig:delta} we observe that 
the aggregate demand was high during the late 1980's, while beginning the 
early 90's, it declined to the bottom in 2000-2001, 
and then, afterward turned up.
It is broadly consistent with changes in the growth rate during the period. 

\figp{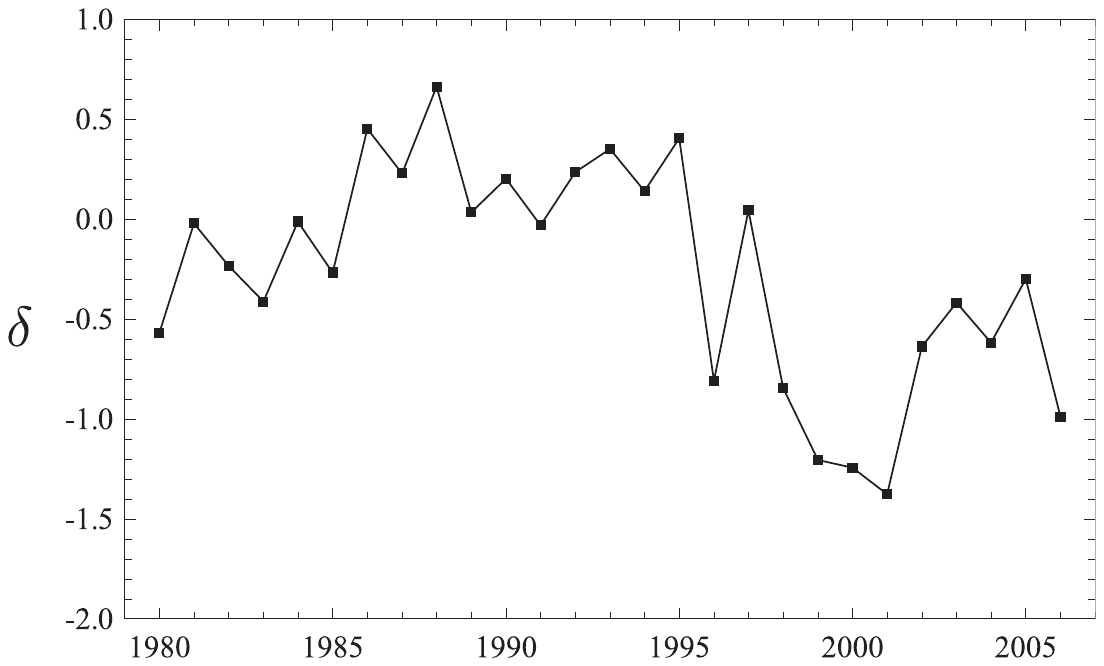}{The values of $\delta$ calculated from Eq.(\ref{mumu})
for the Japanese listed firms.}{0.6}{149 299 462 491}

\vspace{30pt}
\noindent\textbf{\large Acknowledgements}

The authors would like to thank Dr.\ Y.\ Ikeda (Hitachi 
Research Institute) and Dr.\ W.\ Souma (NiCT/ATR CIS)
for their support and discussions.
Part of the research by H.A., H.I., and Y.F.
was supported by a grant from Hitachi Research Institute,
while part of H.Y.'s research was supported by RIETI. 
The authors are grateful to Yukawa Institute for Theoretical Physics for
allowing us the use of their computing facility.

%

\end{document}